\documentclass[fleqn,10pt]{SelfArx} 

\usepackage[english]{babel} 
\usepackage[final]{pdfpages}

\usepackage[T1]{fontenc}
\usepackage[utf8]{inputenc}
\usepackage{authblk}

\usepackage{graphicx,epsfig}

\usepackage{times}
\usepackage{graphics,dcolumn,bm,fleqn,float}
\usepackage{amssymb,amsmath,multirow,rotate,color}
\usepackage{comment} 
\usepackage{cite} 

\definecolor{color1}{RGB}{0,0,90} 
\definecolor{color2}{RGB}{0,20,20} 

\usepackage{hyperref} 
\hypersetup{hidelinks,colorlinks,breaklinks=true,urlcolor=color2,citecolor=color1,linkcolor=color1,bookmarksopen=false,pdftitle={Title},pdfauthor={Author}}

\JournalInfo{~} 
\Archive{} 

\PaperTitle{The joint influence of competition and mutualism on the biodiversity of mutualistic ecosystems} 

\Authors{\normalsize Carlos Gracia-L\'azaro\textsuperscript{1}, Laura Hern\'andez\textsuperscript{2}, 
Javier Borge-Holthoefer\textsuperscript{3}, and Yamir Moreno\textsuperscript{1,4,5}*}

\affiliation{\textsuperscript{1}\textit{Institute for Biocomputation and Physics of Complex Systems (BIFI), University of Zaragoza, Spain.}}
\affiliation{\textsuperscript{2}\textit{Laboratoire de Physique Th\'eorique et Mod\'elisation, UMR CNRS, Universit\'e de Cergy-Pontoise, 2 Avenue Adolphe Chauvin, F-95302, Cergy-Pontoise Cedex, France.}}
\affiliation{\textsuperscript{3}\textit{Internet Interdisciplinary Institute (IN3), Universitat Oberta de Catalunya, Spain.}}
\affiliation{\textsuperscript{4}\textit{Department of Theoretical Physics, Faculty of Sciences, University of Zaragoza, Spain.}}
\affiliation{\textsuperscript{5}\textit{Complex Networks and Systems Lagrange Lab, Institute for Scientific Interchange, Turin, Italy.}}
\affiliation{*To whom correspondence should be addressed: yamir.moreno@gmail.com.} 

\newcommand{\keywordname}{Keywords} 

\Abstract{Relations among species in ecosystems can be represented by complex networks where both negative (competition) and positive (mutualism) interactions are concurrently present. Recently, it has been shown that many ecosystems can be cast into mutualistic networks, and that nestedness reduces effective inter-species competition, thus facilitating mutually beneficial interactions and increasing the number of coexisting species or the biodiversity. However, current approaches neglect the structure of inter-species competition by adopting a mean-field perspective that does not deal with competitive interactions properly. Here, we introduce a framework based on the concept of multilayer networks, which naturally accounts for both mutualism and competition. Hence, we abandon the mean field hypothesis and show, through a dynamical population model and numerical simulations, that there is an intricate relation between competition and mutualism. Specifically, we show that when all interactions are taken into account, mutualism does not have the same consequences on the evolution of specialist and generalist species. This leads to a non-trivial profile of biodiversity in the parameter space of competition and mutualism. Our findings emphasize how the simultaneous consideration of positive and negative interactions can contribute to our understanding of the delicate trade-offs between topology and biodiversity in ecosystems and call for a reconsideration of previous findings in theoretical ecology, as they may affect the structural and dynamical stability of mutualistic systems.}

\begin{document}
\flushbottom 

\maketitle 

\thispagestyle{empty} 

Ecosystems research, and particularly the study of mutualistic communities (like plant and pollinators or plant and seed dispersers), has recently witnessed a major conceptual leap. This has been possible mainly due to two factors: the implementation of more complex dynamical population models that have allowed to go beyond the linear random matrix interaction model of May\cite{may}; and the introduction of a networked perspective that uncovered the consequences of certain structural arrangements at the system level\cite{okuyama,thebault,olesen2007modularity,bastolla2009architecture}, with nestedness at the forefront \cite{jordano2003invariant,bascompte2003nested,bascompte2006asymmetric,fortuna2010nestedness}. Nestedness is a widespread property of mutualistic ecosystems. It allows to quantify key interactions in these systems, particularly the so-called mutualistic interactions, where specialist and generalist species of two guilds (having a small and a large number of inter-guild interactions, respectively, see Figure 1a) facilitate their mutual coexistence. The structure and the dynamics (as given by the persistence or biodiversity of species) of these ecosystems have been shown to be intimately connected\cite{bastolla2009architecture}. 

Additionally, nestedness has emerged as a rather common mesoscale pattern in complex bipartite networks, suggesting that other systems, beyond ecological communities\cite{may2008complex,saavedra2011strong,kamilar2014cultural,borge2015nested,bustos2012,ermann2012}, might follow a similar competition-minimization principle\cite{bastolla2009architecture}. However, both a bipartite representation and the nestedness of the system fail to incorporate intra-guild (competitive) links --so far accounted for within a mean-field description--, precluding current research to develop an even more realistic framework. Although some alternative representations have been recently suggested\cite{lafferty2006parasites,kefi2015network,scotti2013social}, they still fall short when it comes to inspect the feedback between structure and dynamics of mutualistic systems. In parallel, recent advances in data collection suggest that considering actual (rather than probabilistic) information on intra-guild connectivity will be feasible soon\cite{kefi}. It is then imperative to properly deal with both positive (mutualistic) and negative (competitive) interactions in a way that naturally allows to plug in dynamical population models. To this end, here we capitalize on the recently developed framework of multilayer networks\cite{multilayer}, and show that it is possible to encode within a unique topological representation,  {\em both} kinds of interactions. In turn, this allows us to consider, through analytical and numerical results, how the biodiversity of the system varies as a function of the intensities of mutualism and competition in the system.

\onecolumn

In order to build the multilayer representation of a mutualistic ecosystem, we exploit the implicit information contained in the bipartite matrix: the projections onto the subspace of species A (animals) and species P (plants) yield two hidden weighted networks that reveal the existent intra-guild interactions, see Figure 1B, which are encoded in the adjacency matrices associated to each layer.  These topological relations can be used to model inter-species competition beyond the mean-field approach, as they explicitly take into account the actual architecture of intra-layer interactions. Note that in this representation, mutualistic interactions are accounted for by the inter-layer connections between elements of A and P, see Figure 1b. We next investigate the influence of the network structure on the persistence of species of the mutualistic ecosystem. The main point of interest is whether the competitive interactions, as given by the multilayer topology, actually convey substantive dynamical changes. To this end, we implement a population dynamical model that capitalizes on the one introduced by Bastolla {\em et al.} \cite{bastolla2009architecture}, but constrained by the multilayer architecture. We study the variation of biodiversity (e.g., the number of species at the steady state of the population dynamics) on a diverse set of real ecological communities\cite{Herrera-M_PL_016,Kato-M_PL_044,Clements-M_PL_005,Kakutami-M_PL_054,Dupont-M_PL_048,Kato-M_PL_056} (see also the Supplementary Information), for a wide range of parameters of the model.

\begin{figure}[tb!]
\begin{center}
\includegraphics[width=0.7\columnwidth]{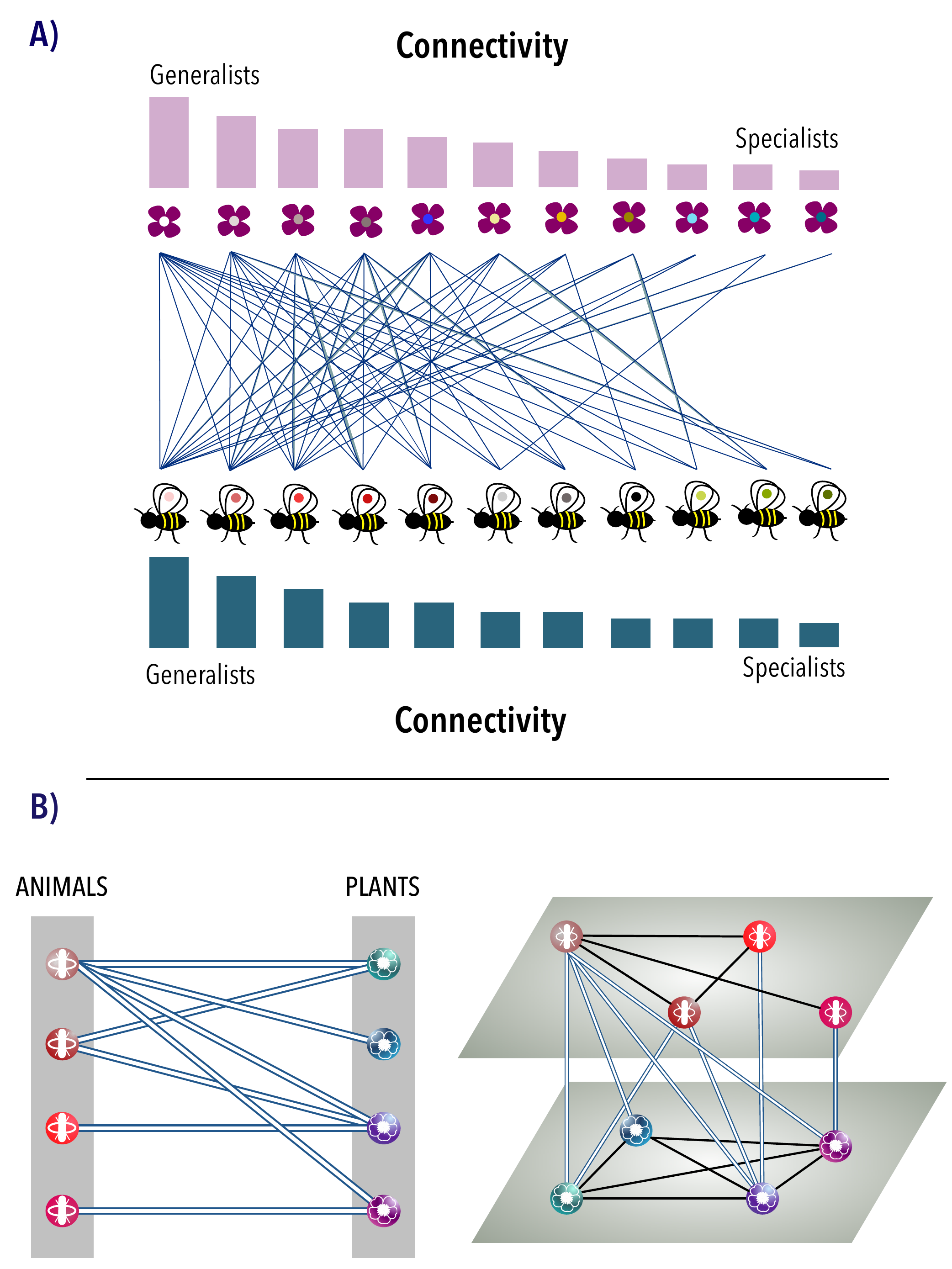}
\vspace{15pt}
\caption{{\bf Multilayer mutualistic network.} Panel a illustrates a mutualistic system made up by plant and animal species. In this representation, mutualistic interactions are given by the inter-connections among the elements of a bipartite graph. Generalists have a higher connectivity than specialists. This representation does not account for intra-guild interactions properly. In panel b, we illustrate the multilayer approach proposed here for an ecosystem of plants and animals that consists of 4 species of each guild. In this framework, each layer represents one guild and an intra-layer link exists whenever two species of the same guild share the same species of the other guild. These links represent the  competition among species of the same guild
leading, in general, to weighted networks. These two layers are coupled by the mutualistic interactions
given by the bipartite graph.}
\label{FigureArtWork}
\end{center}
\end{figure}

Let's assume that the mutualistic community consists of $N^P$ species of plants and $N^A$ species of animals (pollinators or seed-dispersers); the biodiversity is denoted by $N=N^P+N^A$. We denote by $s^P_i$ the abundance of the plant species $i$, and by $\alpha_i^P$ its intrinsic growth rate. Similarly, animals' parameters and variables are represented by the superscript $A$. 
The mutualistic relationships (\textit{i.e.}, the inter-layer connections) are given by a bipartite $N^P \times N^A$ matrix, $K$, with $K_{ik}=1$ if animal species $k$ pollinates the plant species $i$, and  $K_{ik}=0$ otherwise. The biomass of the pollinators of a given plant species $i$ is thus $M^P_i=\sum_{k\in A} K_{ik} s_k^A$. On the other hand, the intra-layer relationships represent the resources that are shared by species of the same guild. Therefore, the biomass of the pollinators shared by two plant species $i,j$ is $W^P_{ij}=\sum_{k\in A} K_{ik} K_{jk} s_k^A$. Finally, the relative abundance of a given plant $i$ evolves according to:

\begin{center}
\begin{eqnarray} \label{dynamicsEquationPlants}
\frac{1}{s_i^P} \frac{d s_i^P}{d t} =
  \alpha_i^P
-\beta_{i}^P s_i^P 
-\beta_0^P \frac{\sum_{j\in P, i\neq j} s_j^P W^P_{ij}}{M^P_i} 
+\gamma_0^P \frac{M^P_i}{ 1+h^P \gamma_0^P M^P_i }\;\; .
\end{eqnarray}
\end{center}

The first term of this equation represents the intrinsic  growth of species $i$  without considering saturation and the second term refers to the intra-specific competition term (saturation), which can be interpreted in terms of a carrying capacity in the absence of competing species. The third term of Eq.~(\ref{dynamicsEquationPlants}) accounts for the intra-guild inter-specific competition. Here, the competition between two plant species ($i,j$) is weighted according to the incidence of the biomass of shared pollinators, $W^P_{ij}$, in the biomass of pollinators of each plant species, $M^P_i, M^P_j$. Lastly, the fourth term in Eq.~(\ref{dynamicsEquationPlants}) gives the contribution of mutualism to the abundance of plant species $i$, $h^P $ being the Holling term that imposes a limit to the mutualistic effect. The intensities of competition and mutualism $\beta_0$ and $\gamma_0$ respectively, constitute the parameter space that we investigate. The corresponding equation for the abundance of pollinators is equivalent to Eq.~(\ref{dynamicsEquationPlants}) but interchanging superscripts $P$ by $A$ and vice versa (see Supplementary Information).

We numerically solved the system of equations describing the abundances of plants and animals (see Methods). Figure 2 compares results obtained for the system's biodiversity when the single (panels a and c) and multilayer (panels b and d) population models are implemented in two real networks (M\_PL\_044\cite{Kato-M_PL_044} in panels a and b; and M\_PL\_048\cite{Dupont-M_PL_048} in panels c and d) as a function of the intensities of mutualism and competition. The single layer population model corresponds to the mean-field version of the competition term\cite{bastolla2009architecture} while the mutualistic term does contain the information about the structure of the real network. On the contrary, the multilayer setup naturally deals with the structure of the competitive interactions. As seen in the figure (see also the Supplementary Information for results corresponding to other real networks), the biodiversity of the system is greatly affected by intra-guild competition. In the mean-field approach, where all species compete on equal grounds with all the others of the same guild, the persistence of biodiversity in the real systems does not depend on $\gamma_0$. Conversely, when the structure of the real network is introduced in the inter-species competition term, the region of structural stability depends, non-trivially, on both parameters, $\gamma_0$ and $\beta_0$. 

The  results shown in Figure 2 may appear counterintuitive. Indeed, if mutualistic interactions were to reduce effective competition and increase biodiversity\cite{bastolla2009architecture}, one should expect that the boundary separating the region where all species survive (coded in red in Fig. 2) from the one where biodiversity diminishes (coded in blue in Fig. 2) would behave as a monotonous growing curve, which is clearly not the case. Figure 3 illustrates this point. As expected, the biodiversity is a decreasing function of the competition parameter ($\beta_0$) in both settings when the mutualism intensity ($\gamma_0$) is fixed (panels c and d). Note, additionally, that in the mean-field case, the persistence is independent of $\gamma_0$, as also shown in Figure 2.Remarkably enough, when the
competition term takes into account the actual interactions in
the multilayer frame, the same decreasing behavior for the persistence of species is also observed when the intensity of competition is kept constant and the extent of mutualism is varied,
see panel b in Fig. 3, except for very low $\beta_0$ values, where competition may be neglected and
the biodiversity does not vary with mutualism. This paradoxical result is due to the joint action of the mutualistic and the inter-species competition terms, which lead to differences on the way the abundance of generalist and specialist species evolves (see Methods for an heuristic argument and the Supplementary Information). The latter selective influence of both mutualism and competition on specialists' and generalists' species is illustrated in Figure 4, where we represent the relative abundance of the species as a function of the species' connectivity and the intensities of mutualism (top panels, in which the competition is fixed) and competition (bottom panels, in which the mutualism is fixed). The results are clear-cut: it turns out that species with higher degrees remain relatively more abundant than those with lower degrees when there is an increase of the strength of either mutualistic or competitive interactions.

\begin{figure}[tb!]
\centering
\includegraphics[width=0.8\columnwidth]{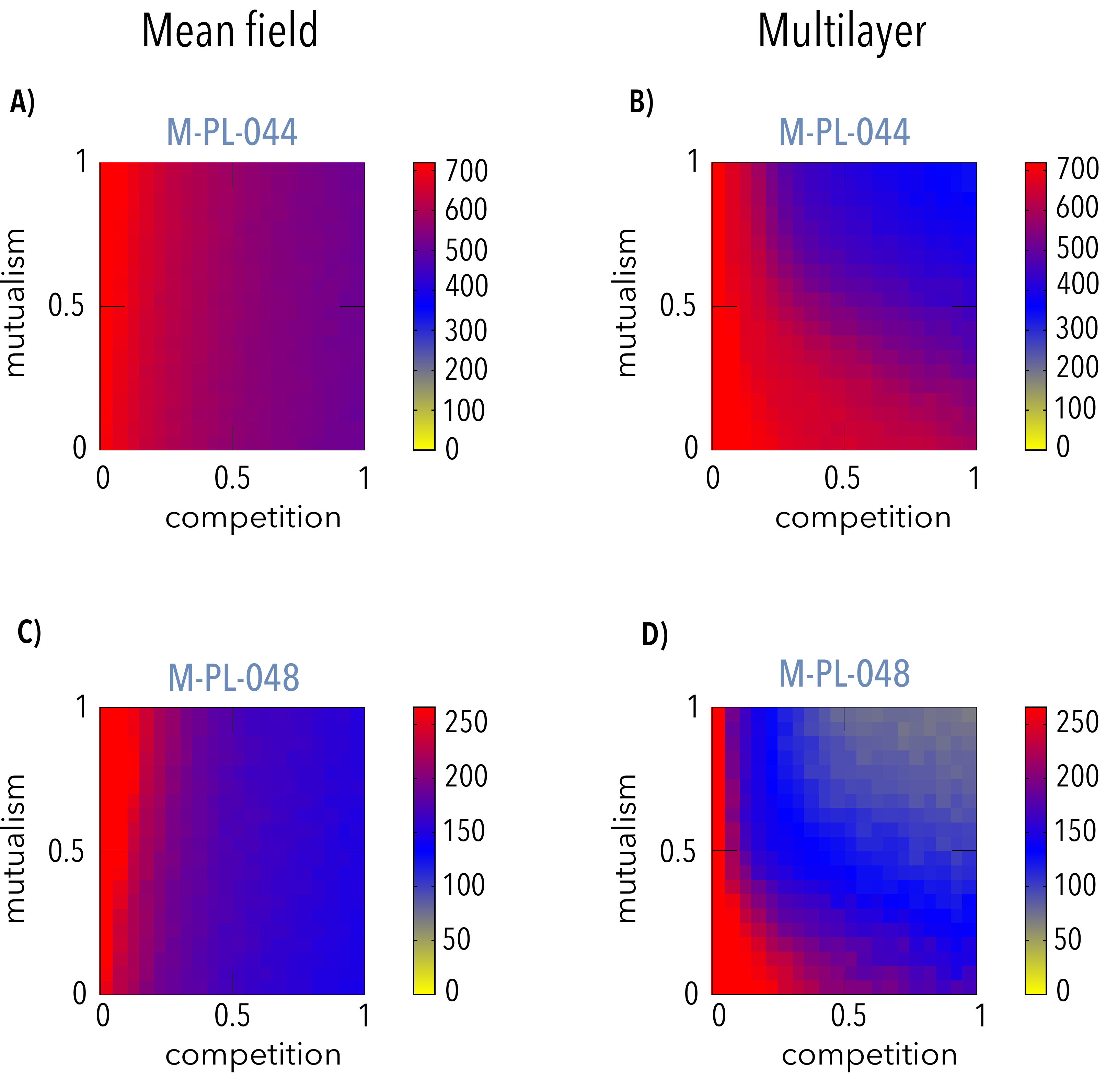}
\vspace{15pt}
\caption{{\bf Biodiversity as a function of the mutualism and intra-guild competition parameters.} We show the results obtained by numerically simulating the dynamical population model when coupled to two real mutualistic networks. The levels of biodiversity are shown as a function of the in-
tensities of mutualism and competition (the maxima being N = 719 for M-PL-044 and
N = 266 for M-PL-048). The color scale represents biodiversity, given by the number, $N$, of  species
present in the steady state. Left panels a and c show the results obtained when the system evolves according to a dynamics where the inter-species competition term is a mean field approximation, as in~\cite{bastolla2009architecture}. Right panels b y d show the results obtained when the competitive interaction are given by the intra-layer links of the multilayer mutualistic network. We show results for two different real networks: M\_PL\_044\cite{Kato-M_PL_044} in top panels a and b; and M\_PL\_048\cite{Dupont-M_PL_048} in bottom panels c and d. The characteristics of these networks as well as more results for other networks are presented in the Supplementary Information. These 
results correspond to the following parameter values: intraspecific competition terms $\beta^P_{i}=\beta^A_{k}=5.0$, growing terms $\alpha^P_i \in (0.9,1.1)$, $\alpha^A_k \in (0.9,1.1)$ and Holling terms $h^P=h^A=0.1$.}
\label{FigureMeanField}
\end{figure}

The results that come out from the approach here adopted unveil the important role played by the network structure of the inter-species competition term on the biodiversity of the system. When this interaction is treated in a mean-field fashion, the biodiversity persists for any intensity of mutualism, as long as the interspecific competition remains under a certain value $-$the frontiers indicating the loss of biodiversity are vertical in panels a and c of Fig. 2. On the contrary, when considering the network structure in the mutualistic \textit{and} in the intra-guild competition term, the region of the parameters space where the biodiversity persists depends non trivially on the intensity of both. In other words, increasing the intensity of mutualism (for a given network) does not necessarily increase biodiversity. Indeed, depending on the intensity of competitive interactions, higher levels of mutualism are detrimental for the survival of the (specialist) species. Therefore, our findings lessen the relevance of the widespread belief that mutualism diminishes competition, thus enhancing biodiversity\cite{bastolla2009architecture} when it comes to analyze the persistence of species in mutualistic systems. The latter is only roughly valid for weak competition levels, when increasing mutualism is not detrimental for the biodiversity (see Figure 3). 

However, our results do not imply that mutualism is not relevant in order to explain the existence of large complex ecosystems. Instead, the careful treatment of the structure of the interactions provides a better understanding of the subtle trade-offs between competition and mutualism. In Ref.\cite{sugihara2009complex} an indirect mechanism for cooperation via the interaction with a common counterpart was discussed. Here we show, in addition, that the asymmetry of the competition term between a generalist and a specialist, induced by mutualism, favors the generalist species. This explains why when the intensity of the mutualistic interactions increases, biodiversity may diminish through an important loss of specialists species in favor of the increase of the population of the generalist ones. Moreover, we have checked that our conclusions are indeed due to the structure of the inter-species competition term and not to the fact that the intensity of this competition is not homogeneous. That is, one might think that similar results would be obtained in a mean-field treatment but using heterogeneous intensities for the inter-species competition term. To show that the latter is not the case, we simulated a system in which the inter-species interaction is such that each species interacts with all the others but with a non homogeneous intensity drawn from a distribution. The results (details can be found in the Supplementary Information) indicate that this heterogeneity is not enough to reproduce the same patterns of biodiversity obtained when the projected, multilayer network representation is considered.

In summary, we have introduced a multilayer approach to the study of mutualistic ecosystems that allows to properly consider competitive interactions among species. Numerical simulations of a dynamical population model that is coupled to the data-driven multilayer architecture revealed that the role of mutualistic interactions in maintaining the system's biodiversity is not as important and as differentiating as previously thought\cite{bastolla2009architecture,rohr2014structural}. Instead, we have reported that both mutualism and competition play a complementary role. Strikingly, we have shown that, contrary to what one would have expect, when the level of competition is high, the biodiversity of the system is higher for lower mutualistic intensities and indeed, increasing mutualism maybe detrimental for the species persistence. In light of the present results, there are a number of further questions that remain to be explored, including whether there is an optimal value of mutualism (and competition) at which the system maximizes biodiversity and its dependency with the nestedness of the system. Considering the multilayer approach introduced here would helpfully provide more realistic grounds to tackle these and related challenges.

\begin{figure}[tb!]
\centering
\includegraphics[width=0.8\columnwidth]{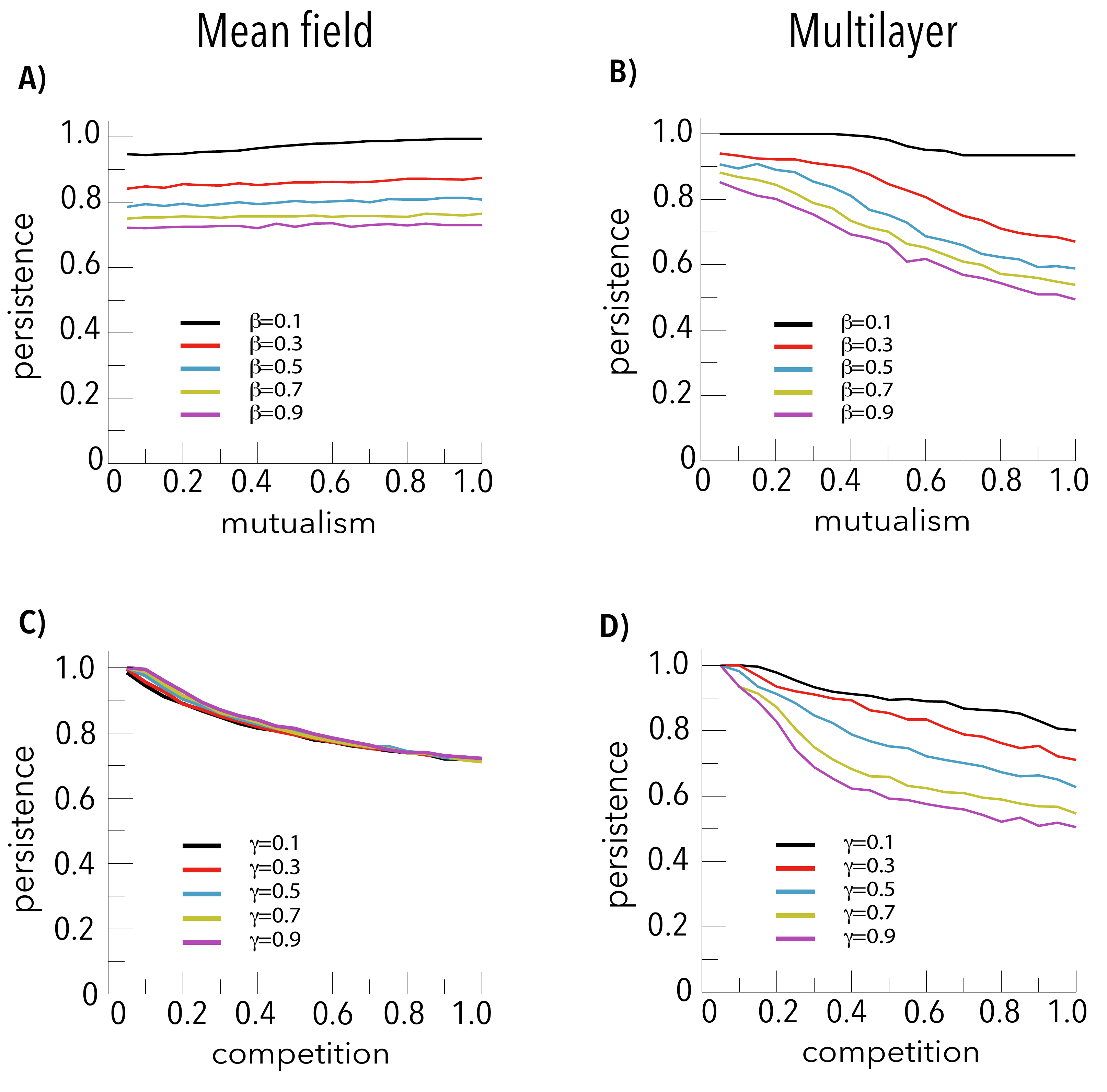}
\vspace{15pt}
\caption{{\bf Persistence of biodiversity.} Top panels a and b show the fraction of species 
in the steady state as
a function of the intensity of mutualism for different constant values of the competition intensity
$\beta_0$ . Complementary, bottom pannels (c and c), show the persistence of biodiversity as a
function of the competition intensity, for different values of fixed mutualism, $\gamma_0$. Left
panels a and c correspond to the bipartite representation of the mutualistic systems, in which
the competitive interactions are given by a mean-field approximation; whereas right panels
b and d show
results obtained when the dynamical population model is constrained by a multilayer
network, thus properly accounting for the competitive interactions.}
\label{FigureFixedBetaGamma}
\end{figure}

\begin{figure}[tb!]
\centering
\includegraphics[width=0.8\columnwidth]{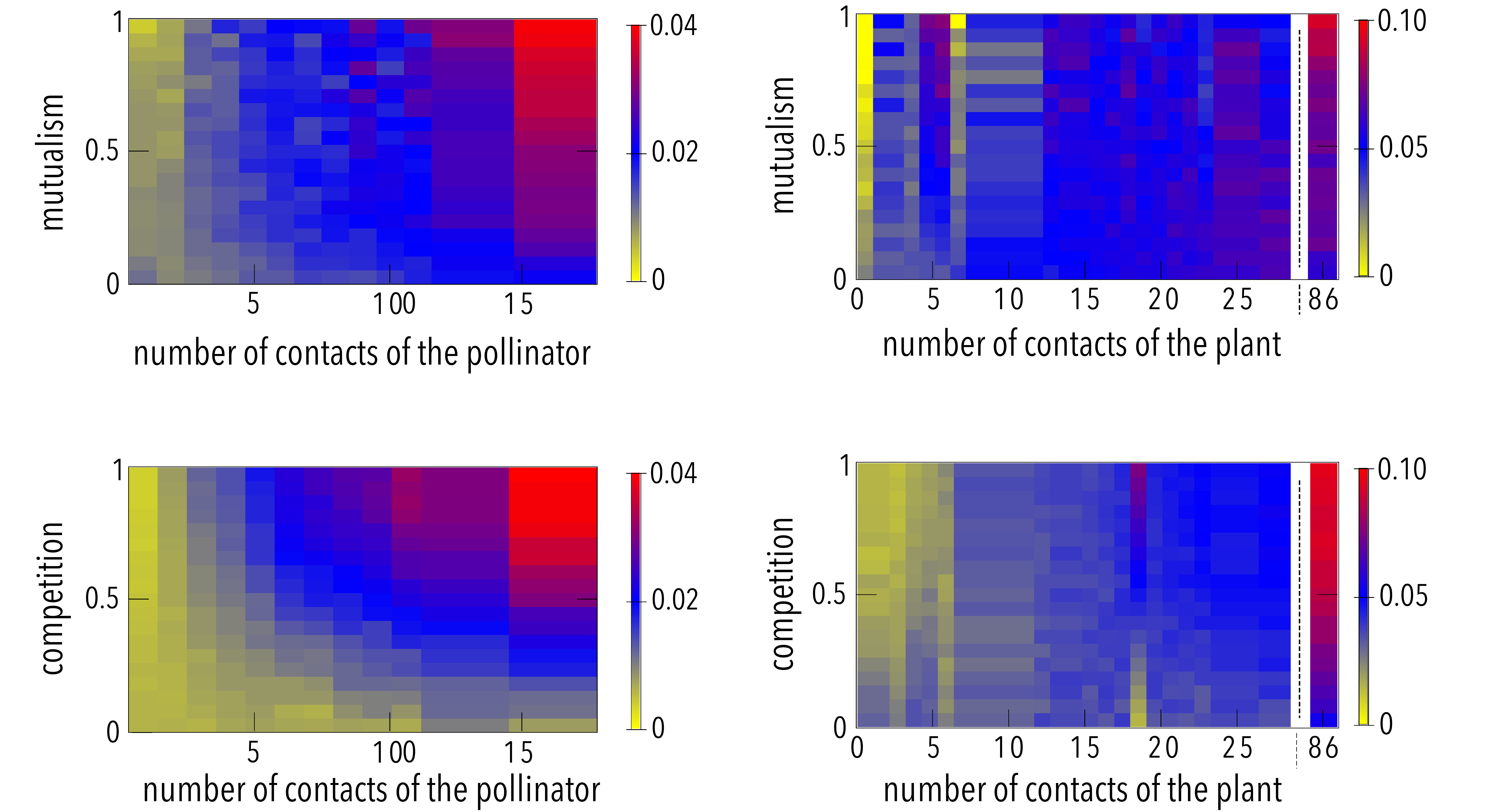}
\vspace{15pt}
\caption{{\bf Relative abundance of the species according to their connectivity, in the multilayer model.} 
The results shown here correspond to the evolution of the system according to equations (\ref{dynamicsEquationPlants}) and (\ref{dynamicsEquationAnimals}),  with the interaction terms constrained by the real network M-PL-016 (left pannels: pollinator species; right pannels: plant species). Top panels: The competition intensity
is fixed to $\beta_0 = 0.25$, the color scale represents the relative abundance of the different
species, as a function of the mutualism parameter, $\gamma_0$ , and inter-layer connectivity. Bot-
tom panels: The mutualism interaction is fixed to $\gamma_0 = 0.25$,the color scale represents the
relative abundance of the different species, as a function of the competition parameter $\beta_0$
and interlayer connectivity. The gaps on right panels correspond to non existing values
of the plant’s degree. Animals (left panels) and plants (right panels) have been ranked in
ascending order of interlayer degree. The rest of parameters are the same as in Figure 1.}
\label{FigureFrequencyDegree}
\end{figure}

\section*{Methods}

\paragraph{Equation for species A.} The dynamical equation that describes the evolution of the abundance of animal species $i$ is equivalent to Eq.\ (\ref{dynamicsEquationPlants}) in the main text, i.e.,
\begin{eqnarray} \label{dynamicsEquationAnimals}
\frac{1}{s_k^A} \frac{d s_k^A}{d t} =
\alpha_k^A
-\beta_{k}^A s_k^A 
-\beta_0^A \frac{\sum_{l\in A, k\neq l} s_l^A W^A_{kl}}{M^A_k}
+\gamma_0^A\frac{M^A_k}{ 1+h^A \gamma_0^A M^A_k } .
\end{eqnarray}

\paragraph{Numerical Simulations of the Model.} We numerically solve the system of equations (\ref{dynamicsEquationPlants}) and (\ref{dynamicsEquationAnimals}), using the matrices $K_{ik}$ that correspond to different real systems (see also Supplementary Information), namely, M\_PL\_016\cite{Herrera-M_PL_016}, M\_PL\_044\cite{Kato-M_PL_044}, M\_PL\_005\cite{Clements-M_PL_005}, M\_PL\_054\cite{Kakutami-M_PL_054}, M\_PL\_048\cite{Dupont-M_PL_048} and M\_PL\_056\cite{Kato-M_PL_056}. Each simulation starts from random initial conditions of the relative abundances. Following Ref.\cite{bastolla2009architecture,rohr2014structural} we take the values of $\alpha_i^{P,A}$ from a uniform distribution in the interval  [0.9, 1.1],  the intra-species competition is fixed to $\beta_{j}^{P,A }=5 $ and the Holling term is  $h^{P,A} = 0.1$. With these parameters, we study the system varying the value of the intensity of the inter-species competition and mutualistic terms, $\beta_0^{P(A)}$ and $\gamma_0^{(P)A}$, respectively. For simplicity, we assumed that all the intervening parameters take the same values for plants and animals. Finally, the system is considered to have achieved equilibrium when all the species' frequencies remain constant. A species is considered to have gone extinct when its relative abundance is lower than $10^{-9}$.

\paragraph{Evolution of the abundance of generalist and specialist species.} In order to understand how the population dynamics of generalists and specialists is affected by mutualistic and competitive terms, let us first consider the effect of the mutualist term (the fourth term in Eqs.~(\ref{dynamicsEquationPlants}) and~(\ref{dynamicsEquationAnimals})). This term reads: $\frac{M^P_i}{1+\delta M^P_i}$, where $\delta=h \gamma_0$. As $M^P_i$ is larger for generalists than for specialists, the increasing rate of the former is stronger than that of the latter. In other words, this term favors the increase of generalists species with respect to specialists ones. The analysis of the inter-species competition term (the third term of  Eqs.~(\ref{dynamicsEquationPlants}) and~(\ref{dynamicsEquationAnimals})) is less straightforward. Let us compare the behavior of this term for a generalist and a specialist  plant species, of relative abundances $s_1^P$ and  $s_2^P$, respectively, in the particular case where the specialist interacts with only one animal species, $l$, of abundance $s_l^A$. This means that $K_{2k}=\delta_{kl}$ and $M^P_2=s^A_l$. In this extreme case, it is  very easy to see that the competing term that enters in Eq.~(\ref{dynamicsEquationPlants}) for species $2$ reads:
\begin{equation}
C_2= \beta_0^P  \lbrace s_1^P + \sum_{j\in P, j \ne 1,2} s_j^P  K_{jl} \rbrace
 \end{equation}
\noindent where the first term corresponds to its competition with the generalist species $1$, and the second term stands for the competition with all the other plant species that share the same pollinator, $l$, of abundance $s_l^A$. The corresponding competition term for the generalist $1$ is 
\begin{equation}
\label{compet_generalist}
C_1 = \beta_0^P \sum_{j\in P, j\ne 1} s_j^P \lbrace \frac{\sum_{k\in A} K_{1k} K_{jk} s_k^A}{\sum_{k\in A} K_{1k} s_k^A}\rbrace
\end{equation}
\noindent where the factor between braces is $\alpha_j \le 1$. The latter expression can be rewritten as:
\begin{equation}
\label{compet_generalist_short}
C_1= \beta_0^P (s_2^P \alpha_2 +  \sum_{j\in P, j \ne 1,2} s_j^P \alpha_j)
\end{equation}
The second term of eq.~\ref{compet_generalist_short}, which may include other generalist plants (those that grow faster with $\gamma_0$), is reduced by the factors $\alpha_j \le 1$, and in general $C_2 > C_1$. Therefore, we can then expect that this unbalance in the corresponding competition terms becomes a supplementary advantage for the generalists, thus reinforcing their growth rate. A similar analysis to estimate the relative importance of the competition term can be done using the properties of projected matrices (see the Supplementary Information).

\newpage

\onecolumn

\section*{SUPPLEMENTARY INFORMATION}

\subsection*{Multilayer Networks}

A graph (\textit{i.e.}, a single-layer network) is a tuple $G = (V , E)$, where $V$ stands for the set of nodes
and $E\subset V\times V$
is the set of edges that connect pairs of nodes. The edges of the graph induce a binary relation on $V$ that is called the adjacency relation of $G$. This relation can be represented through the adjacency matrix $\mathbf{A}$, where  $A_{i,j}$ indicates the number of
links from node $i$ to node $j$.

As a particular case, a bipartite graph is a graph whose nodes can be divided into two disjoint sets $V_1$ and $V_2$
($V=V_1 \cup V_2$,
$V_1\cap V_2=\emptyset$) such that every edge connects a node in $V_1$ to one in $V_2$. The adjacency matrix of a
bipartite graph has the form of a block matrix:

\begin{equation} 
\mathbf{A} =
   \begin{bmatrix}
      \mathbf{O}_{n \times n} & \mathbf{\tilde{A}}_{n \times m}\\
      (\mathbf{\tilde{A}}^{T})_{m \times n} & \mathbf{O}_{m \times m}
   \end{bmatrix}\;\;, 
   \label{bipartiteFromula}
\end{equation}

\noindent where $\mathbf{O}$ represents the $n\times n$ null matrix, $O_{ij}=0$.

A multilayer network is a data structure made of multiple layers, where each layer is a single-layer
network. Formally, a multilayer network is a tuple $M = (\mathcal{G},C)$ where
$\mathcal{G} = \lbrace G_\alpha; \alpha \in \lbrace 1 , . . . , n \rbrace\rbrace$ is a family of
graphs $G_\alpha = (V_\alpha, E_\alpha)$ (the layers), and 
$C\in V_\alpha \times V_\beta$ ($\alpha ,\beta = 1,2,\ldots,n; \alpha \neq \beta$)
is the set of edges between nodes of different layers. 

In order to build a more complete network containing the interactions among species of different guilds as well as intra-guild interactions, we can further exploit the information contained in the bipartite network of mutualistic interactions and consider the projections of the mutualistic interactions onto the guilds of plants ($P$) and animals ($A$). This set-up gives us
a multilayer network where the guilds constitute the layers, the mutualistic interactions provide the interlayer links and their projections account for the intra-layer links, which in turn represent the number
of pollinators (\textit{resp.}, plants) shared by the corresponding plants (\textit{resp.}, pollinators). Accordingly, the resulting multilayer network consists of two layers ($V_P, V_A$), the set of inter-layer links ($C\in V_P \times V_A$), and two sets of intra-layer links ($E_P\in V_P \times V_P$, $E_A\in V_A \times V_A$). 

For the sake of clarity, here we adopt the notation commonly used in the biological literature. The mutualistic
plant-pollinator interactions (\textit{i.e.}, the inter-layer connections) are given by a
bipartite $N^P \times N^A$ matrix, $K$, with $K_{ik}=1$ if animal species $k$ pollinates the
plant species $i$, and  $K_{ik}=0$ otherwise. The projection of this matrix onto
the sets of plants is given by:

\begin{equation} \label{projectionPlants}
V^P_{ij}=\sum_{k\in A} K_{ik} K_{jk} \;\;,
\end{equation}

\noindent where $V^P_{ij}$ represents the number of pollinators shared by plant species $i$ and $j$.

In our dynamical equations, we consider the resources that are shared by species of the same
guild. According to Eq. \ref{projectionPlants}, the biomass of the pollinators shared by
plant species $i,j$ is given by:

\begin{equation} \label{sharedBiomass}
W^P_{ij}=\sum_{k\in A} K_{ik} K_{jk} s_k^A \;\;.
\end{equation}

Equivalently, the respective formulas for pollinators are obtained by interchanging the $P$ and $A$ superscripts and vice versa, that is,

\begin{equation} \label{projectionAnimals}
V^A_{kl}=\sum_{i\in P} K_{ik} K_{il} \;\;,
\end{equation}

\begin{equation} \label{sharedBiomassAnimals}
W^A_{kl}=\sum_{i\in P} K_{ik} K_{il} s_k^P \;\;.
\end{equation}

Table \ref{tab1} shows the main features (number of species, plants and animals),
the location of the biotope and the reference of the ecological networks used to numerically solve the differential equations.

\begin{table}[t]
\begin{center}
\begin{tabular}{cccccc}
network & species & plants & pollinators & location & reference \\
\hline\hline
M-PL-016 & 205 & 26 & 179 & Do\~nana Nat. Park, Spain & \cite{Herrera-M_PL_016}\\
\hline
M-PL-044 & 719 & 110 & 609 & Amami-Ohsima Island, Japan & \cite{{Kato-M_PL_044}}\\
\hline
M-PL-005 & 371 & 96 & 205 & Pikes Peak, Colorado, USA & \cite{Clements-M_PL_005}\\
\hline
M-PL-054 & 431 & 113 & 318 & Kyoto, Japan & \cite{Kakutami-M_PL_054} \\
\hline
M-PL-048 & 266 & 30 & 236 & Denmark & \cite{Dupont-M_PL_048} \\
\hline
M-PL-056 & 456 & 91 & 365 & Mt. Kushigata, Japan & \cite{Kato-M_PL_056}\\
\hline\hline
\end{tabular} \caption{{\bf Ecological netwoks.} Values for
the number of species, plants and animals, location of
the study and reference of the plant-pollinator
ecological networks used in this study.} \label{tab1}

\end{center} 
\end{table}

\subsection*{Distribution of the competitive term in the mean field approach}

Here we explore an alternative formulation of the mean-field treatment. The idea is to explore whether the results obtained are truly due to the new topological constraints (as given by the multilayer network) or if it would be equivalent to introduce the heterogeneity via the intensity of the inter-species competition term. We show below that this is not the case, that is, that the observed trade-off between competition and mutualism when the full network structure is considered in both mutualistic and competitive interactions (Figure 2 of the main text) is not present when a mean-field approximation is adopted for the competition (see also Figure 2 of the main text). Specifically, we have studied the dynamics \cite{bastolla2009architecture} with a heterogeneous distribution of the competitive strength, that is,
the inter-species competition terms $\beta_{ij}$ are taken from an uniform distribution
$(0.05\beta_0,1.95\beta_0)$. 

\begin{figure}[ht]
\centering
\includegraphics[width=0.75\columnwidth]{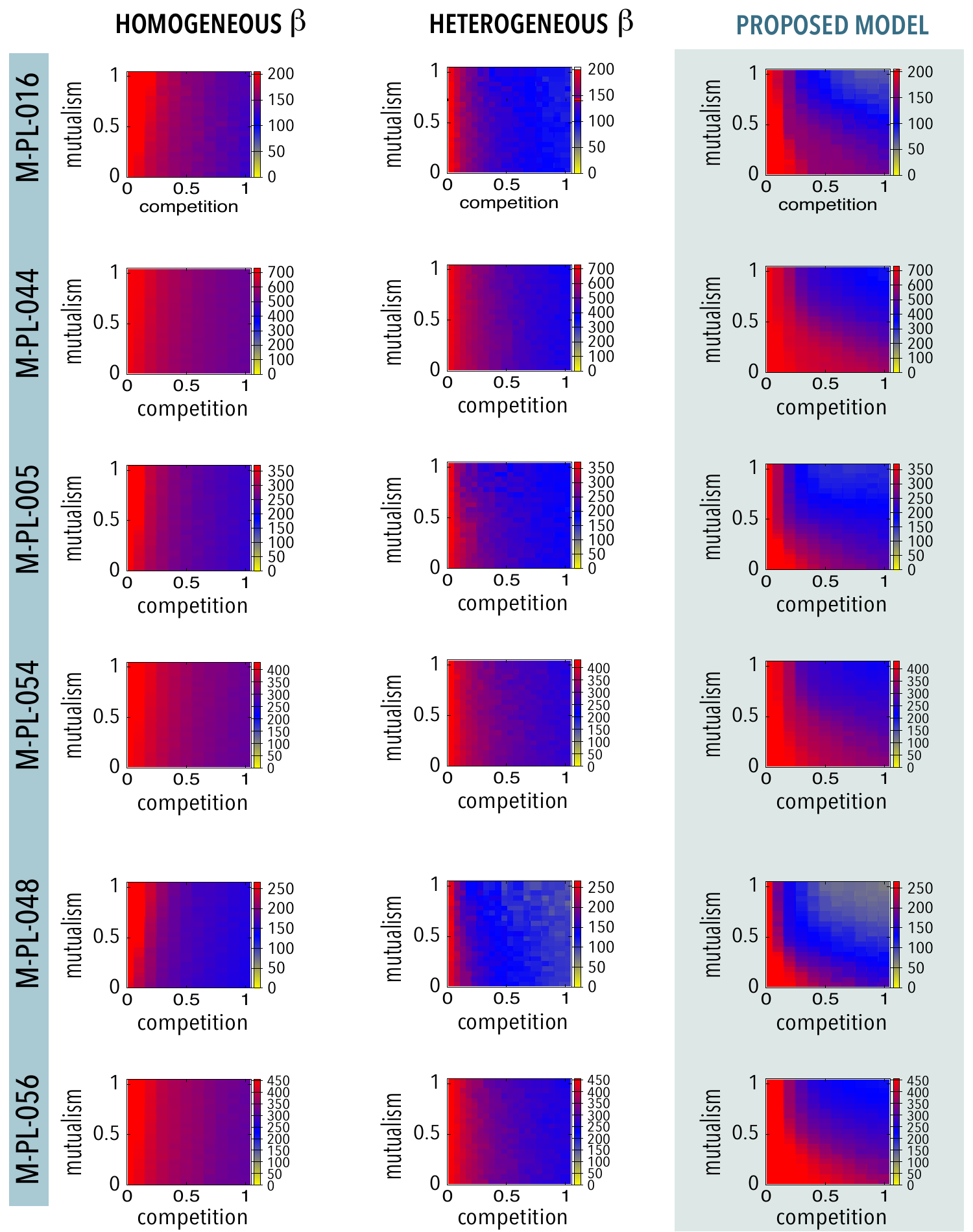}
\vspace{15pt}
\caption{{\bf Biodiversity in mutualistic networks under different treatments of the competition term.}
The color-coded values in the figure represent the number $N$ of final species as a function of the inter-species competition and mutualistic terms (\textit{resp}., $\beta_0$ and $\gamma_0$), for different mutualistic networks (rows), after the system has evolved according to the dynamical population model. Left and central panels correspond to the mean field approximation for the competition term, when a homogeneous distribution of the inter-specific competition term $\beta_{ij}=\beta_0$ (left panels) is considered and when
the inter-species competition intensities, $\beta_{ij}$, are taken from a uniform distribution $(0.05\beta_0,1.95\beta_0)$ (central panels). Finally, the right panels stand for the model here presented. The maximum diversity, corresponding to no extinctions, corresponds to $N= 205, 719, 371, 431, 266, 456$ species for networks M-PL-016, M-PL-048, M-PL-005, M-PL-054, M-PL-044 and M-PL-056 respectively. Other values are: intra-specific competition terms $\beta^P_{i}=\beta^A_{k}=5.0$, growing terms $\alpha^P_i \in (0.9,1.1)$, $\alpha^A_k \in (0.9,1.1)$ and Holling terms $h^P=h^A=0.1$.}
\label{homogenousVsHeterogeneousBeta}
\end{figure}

Figure S\ref{homogenousVsHeterogeneousBeta} represents
the species persistence as a function of the inter-species competition and mutualistic terms (\textit{resp}., $\beta_0$ and $\gamma_0$). The rows correspond to different real mutualistic networks as indicated. Left panels show the results of integrating the dynamical equations when the competitive term is the same for all the species $\beta_{ij}=\beta_0$ (given by the horizontal axis). Some
of them correspond to those shown in the main text. Additionally, the central panels correspond to the case in which the intensities of the inter-species competition term are taken from a uniform distribution
$\beta_{ij}\in (0.05\beta_0,1.95\beta_0)$. As it can be seen, a heterogeneous competition implies a lower biodiversity regarding the homogeneous distribution of the competitive term. Finally, the right panels correspond to the model here presented, defined by equations equations (1) and (2) of the main text, and therefore they represent the case in which the competition terms have been weighted proportionally to the shared biomass of pollinators (\textit{resp}., plants). Again, some of them have been introduced
in the main text. A direct comparison of the central and
right panels clearly shows that introducing (to some extent artificially) a degree of heterogeneity in the competitive terms cannot reproduce the results obtained when considering the whole network structure characterizing the competitive interactions. 

\subsection*{Average relative abundance for animals.}

Here we further develop the heuristic argument provided in the Methods section of the main text to explain the effects of the balance between mutualism and inter-specific competition. Let us consider the evolution of the abundance of a plant species $i$,  $s_i^P$. The factor $W^P_{ij}$ of eq. (\ref{sharedBiomass}) involves the abundance of each animal species of the other guild. 
These abundances are all different because for the animal guild, generalists and specialists also grow faster and slower with $\gamma_0$, respectively. We will then consider an average relative abundance for animals $\langle s^A\rangle$. Then the competing term, for plants, reads:

\begin{equation}
C_i \approx \beta_0^P \sum_{j\in P, j \ne i} s_j^P \lbrace\frac{\langle s^A\rangle V_{ij}^P}{\sum_{k\in A} K_{ik} \langle s^A\rangle}\rbrace 
\end{equation}
 
\noindent where $V_{ij}^P$ is the projected matrix on the plant space.

As $V_{ij}^P \leq min(k_i^P,k_j^P)$ we can consider two situations:
\begin{itemize}
\item $i$ is a generalist  so $k_i^P \geq  k_j^P \forall j$ then, the maximum possible contribution to the competition term is :
 $C_{gen}^{MAX} \approx \beta_0^P \sum_{j\in P, j \ne i} ( s_j^P \frac{k_j^P}{k_i^P})$
\item $i$ is a specialist, so $k_i^P  < k_j^P$ then the maximum contribution to the competition term is :
 $C_{spe}^{MAX} \approx \beta_0^P \sum_{j\in P, j \ne i} s_j^P $

\end{itemize}

\noindent which shows that the specialists are, on average, more affected by competition than generalists when mutualism increases. The same reasoning holds symmetrically for the animal's guild.

\end{document}